\pdfoutput=1

\documentclass[11pt]{article}

\usepackage[final]{acl}

\usepackage{times}
\usepackage{latexsym}

\usepackage[T1]{fontenc}

\usepackage[utf8]{inputenc}

\usepackage{microtype}

\usepackage{inconsolata}

\usepackage{graphicx}
\usepackage{tabularx}
\usepackage{booktabs}
\usepackage{enumitem}
\usepackage{tcolorbox}
\usepackage{multirow}

\title{Beyond Single-Audio: Advancing Multi-Audio Processing\\ in Audio Large Language Models}

\author{Yiming Chen$^{\dag}$ \quad Xianghu Yue$^{\dag\thanks{\quad Corresponding author.}}$ \quad Xiaoxue Gao$^{\P}$ \quad Chen Zhang$^{\dag}$ \\ \textbf{Luis Fernando D'Haro}$^{\star}$ \quad \textbf{Robby T. Tan}$^{\ddag, \dag}$ \quad \textbf{Haizhou Li}$^{\natural, \dag}$ \\
        $^\dag$National University of Singapore \quad 
        $^{\natural}$The Chinese University of Hong Kong, Shenzhen \\
        $^{\star}$Universidad Politécnica de Madrid \quad
        $^{\ddag}$ASUS Intelligent Cloud Services\quad \\
        $^{\P}$I2R, Agency for Science, Technology, and Research (A*STAR) \\
        \tt \{yiming.chen,xianghu.yue,chen\_zhang\}@u.nus.edu \\ \tt Gao\_Xiaoxue@i2r.a-star.edu.sg\quad luisfernando.dharo@upm.es  \\
   \tt  robby\_tan@asus.com\quad haizhouli@cuhk.edu.cn \\
}

\begin{document}
\maketitle

\begin{abstract}
Various audio-LLMs (ALLMs) have been explored recently for tackling different audio tasks simultaneously using a single, unified model.
While existing evaluations of ALLMs primarily focus on single-audio tasks, real-world applications often involve processing multiple audio streams simultaneously.
To bridge this gap, we propose the first multi-audio evaluation (MAE) benchmark that consists of 20 datasets from 11 multi-audio tasks encompassing both speech and sound scenarios.
Comprehensive experiments on MAE demonstrate that the existing ALLMs, while being powerful in comprehending primary audio elements in individual audio inputs, struggling to handle multi-audio scenarios.
To this end, we propose a novel multi-audio-LLM (MALLM) to capture audio context among multiple similar audios using discriminative learning on our proposed synthetic data.
The results demonstrate that the proposed MALLM outperforms all baselines and achieves high data efficiency using synthetic data without requiring human annotations.
The proposed MALLM opens the door for ALLMs towards multi-audio processing era and brings us closer to replicating human auditory capabilities in machines.~\footnote{Code is available at \url{github.com/MatthewCYM/MALLM}.}

\end{abstract}

\section{Introduction}
Large language models (LLMs) have become remarkably powerful, driving advancements in various tasks across the field of natural language processing (NLP)~\citep{touvron2023llama,achiam2023gpt,team2023gemini}.
Recent advancements in LLMs have also led to the development of various powerful audio large language models (ALLMs)~\citep{chu2023qwen,huang2024audiogpt,rubenstein2023audiopalm}, which have achieved impressive results on a range of audio tasks, e.g., automatic speech recognition~\cite{hu2024wavllm}, speech synthesis \cite{gao2024emo}, sound event classification~\cite{tang2024salmonn, coavt}.

\begin{figure}[t!]
  \centering
  \includegraphics[width=0.9\columnwidth]{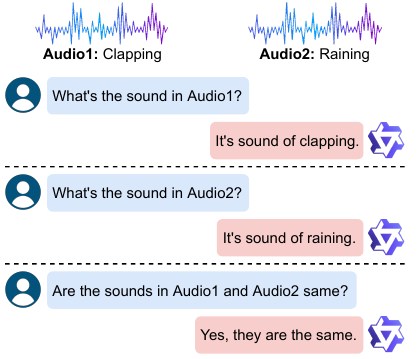} 
  \caption {Motivating Example - Qwen-Audio's responses for single audio inputs and a two-audio input task.}
  \label{fig:motivating-example}
\end{figure}

However, a crucial limitation exists: current ALLM training and evaluation primarily focus on single audio inputs. 
This is a significant drawback, as various real-world applications, e.g., virtual assistants, often require processing multiple audio streams simultaneously.
Additionally, multi-audio processing is essential for effectively implementing few-shot in-context learning, which is a fundamental capability for advanced LLMs. 
Unlike text-based LLMs, which excel at handling multiple texts~\cite{wang-etal-2024-rethinking, mckenna-etal-2023-sources}, and vision LLMs, which have established benchmarks for processing multiple images~\cite{li2024finetuning, huang2023sparkles, zhao2024mmicl, li2023textbind}, the audio field lacks systematic evaluations and benchmarks for multi-audio tasks with ALLMs.
Although several ALLMs~\cite{chu2023qwen, zhan2024anygpt} claim to handle multi-audio contexts, their performance quantification remains unclear. 
This underscores a significant gap in the audio field, in contrast to the vision and language fields, which already possess established evaluations and benchmarks for their respective tasks.

To bridge this gap, we propose the first dedicated multi-audio benchmark for ALLMs, encompassing 11 tasks for both open-ended and close-formed generation across speech and sound domains. 
Comprehensive experiments on 15 ALLMs reveal that existing open-source ALLMs fall short in multi-audio scenarios despite their strong single-audio processing abilities.
For instance, as shown in Fig.~\ref{fig:motivating-example}, while ALLMs can perfectly identify Audio1 as clapping and Audio2 as raining respectively in a single-audio scenario, they struggle to determine whether Audio1 and Audio2 are the same when tasked with understanding the relationship between two audio inputs.
This suggests that current ALLMs are not well-equipped to handle multi-audio tasks, even those that are straightforward for humans. 
This further underlines the need for advancements in ALLMs to handle multiple audio inputs to enhance human-computer interaction.

To this end, we propose an innovative and scalable multi-audio large language model (MALLM) that effectively captures audio contexts essential for reasoning over multiple audios. 
Inspired by the success of discriminative learning~\citep{Clark2020ELECTRA,li2024finetuning,pmlr-v139-jia21b}, we define a challenging discriminative task that trains the model to discover the subtle differences between two similar audio samples.
Furthermore, we introduce a scalable audio pairs synthesis strategy to enable multi-audio processing ability without the need for data collection and human labeling. 
Comprehensive experiments show that MALLM significantly outperforms existing open-source ALLMs under multi-audio scenarios while maintaining competitiveness under single-audio scenarios.

Overall, our contributions include:
\textbf{(1) Novel Evaluation Benchmark:} We propose the first multi-audio benchmark (MAE) for evaluating the multi-audio processing capabilities of ALLMs, encompassing a diverse range of tasks from both the speech and sound domains.
\textbf{(2) Advanced Multi-Audio LLM:} Beyond focusing on the single audio content, our proposed MALLM not only demonstrates remarkable performance across diverse multi-audio tasks but also achieves remarkable data efficiency through an innovative data synthesis strategy.
\textbf{(3) Comprehensive Evaluation:} We conduct a comprehensive evaluation on 15 ALLMs across various tasks, providing a solid foundation for future research.

\section{Related Works}
Recently, numerous studies have integrated audio encoders with pre-trained LLMs to develop multimodal ALLMs, serving as general-purpose task solvers for various audio-input tasks across speech and sound domains. 
Most ALLMs focus on specific audio types, like speech~\citep{zhang-etal-2023-speechgpt,zhan2024anygpt,das2024speechverse}, or sound~\citep{panagopoulou2023x,kong2024audio,moon2023anymal,han2023imagebind}.
A few recent models handle multiple types of audio, showing strong capabilities in universal audio understanding~\citep{gong2023joint,chu2023qwen,tang2024salmonn}.
Meanwhile, to effectively benchmark the advancements in ALLMs, new evaluation benchmarks such as AIR-Bench~\citep{airbench} and Dynamic-SUPERB~\citep{dynamic-superb} have been introduced. 
However, these benchmarks and existing ALLMs primarily focus on tasks involving single audio inputs, largely overlooking scenarios with multiple audio streams.
In contrast, our work introduces the MAE benchmark, specifically designed to assess multi-audio processing capabilities. 
Notably, MAE includes a variety of tasks across diverse scenarios, encompassing both audio and speech, making it suitable for benchmarking a broad spectrum of ALLMs. 
Additionally, we develop the MALLM, the first ALLM specifically tailored for multi-audio tasks, demonstrating significant improvements in processing multiple audio streams while maintaining competitive performance on single-audio tasks.

\section{MAE Benchmark}
\begin{figure*}[t!]
\centering
  \includegraphics[width=0.98\linewidth]{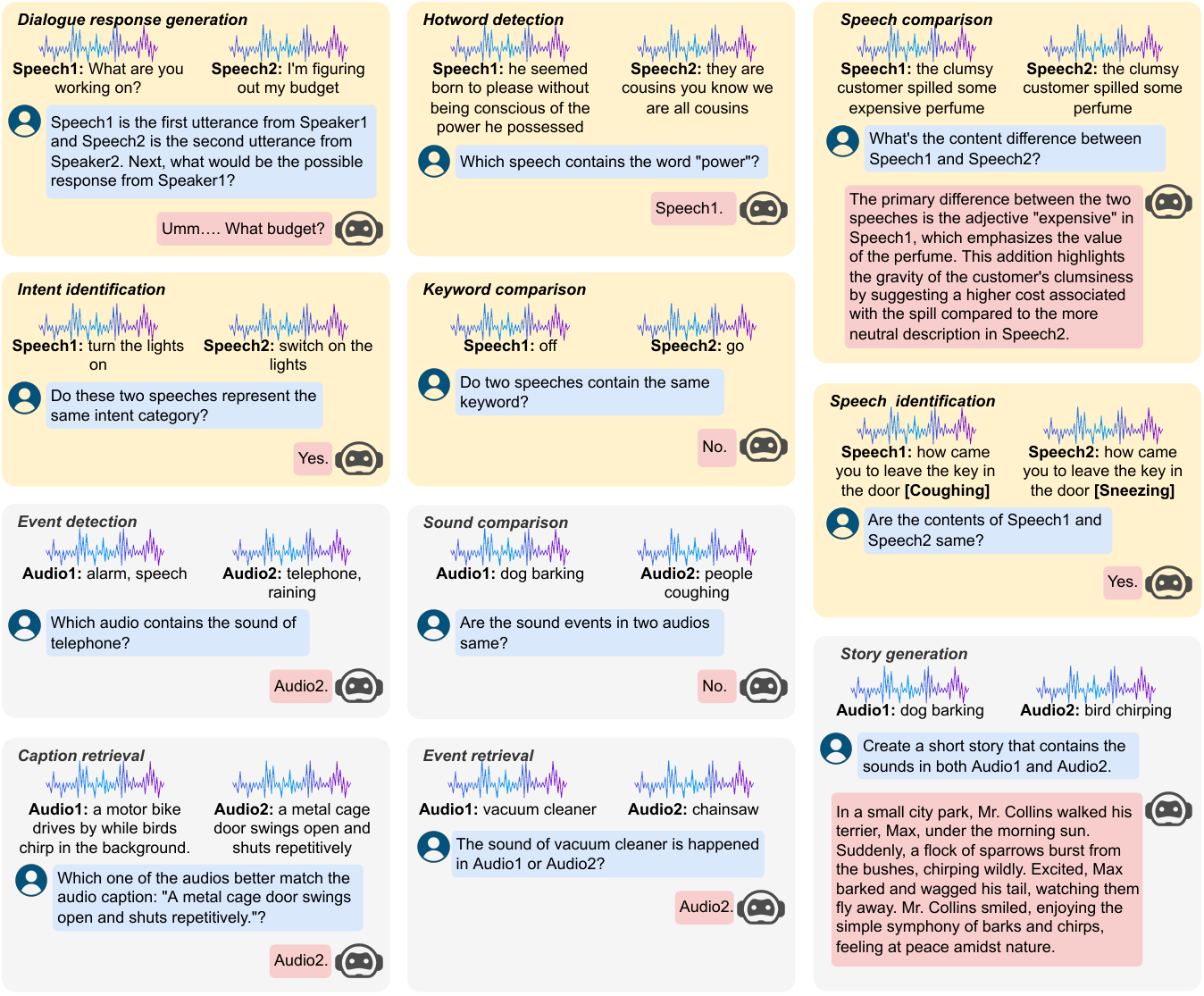} 
  \caption {Overview of proposed MAE benchmark. Yellow blocks show the speech tasks, while grey blocks show the sound tasks. Speech transcriptions, audio labels, input instructions, and expected output responses are given.}
  \label{fig:eval-overview}
\end{figure*}

We propose the first multi-audio benchmark (MAE) for evaluating ALLMs, as illustrated in Fig.~\ref{fig:eval-overview}.
The MAE comprehensively benchmarks the multi-audio processing capabilities of ALLMs by including a wide variety of generation tasks from different fields and scenarios.
It comprises six speech tasks and five sound tasks, covering both open-ended and close-form generation tasks.
Open-ended questions allow models to produce free-form responses without predefined constraints.
Conversely, closed-form questions restrict models to a predetermined set of possible outcomes.
The MAE is automatically constructed using an advanced text-only LLM from various existing single-audio datasets without the need for further human annotation. 
Each sample in the MAE includes a combination of two audio contexts and a task instruction.

To correctly answer the questions from MAE, ALLMs are required to effectively combine the information from both context audios.
Notably, MAE is designed to examine ALLMs' multi-audio processing ability at different levels.
For closed-form generation tasks, we directly derive the ground truth labels from existing single-audio labels to compute metric scores.
Motivated by the recent success of LLM-based text evaluation~\citep{fu2023gptscore}, we utilize a text-only LLM as an automatic evaluator to assign binary good/bad labels to responses generated by ALLMs for open-ended generation tasks (see APPX.~\ref{appx:eval-prompt}).
We summarize the data statistics of the MAE in Tab.~\ref{tab:data-stats}.
Next, we introduce the construction process for each task.

\begin{table*}[t]
\centering
\resizebox{0.98\textwidth}{!}{ 
\begin{tabular}{llcccc}
\toprule
\textbf{Type}& \textbf{Task} & \textbf{Category}& \textbf{Source} & \textbf{\# Samples} & \textbf{Metric} \\
\midrule
\multirow{11}{*}{Speech} & \multirow{2}{*}{Speech comparison}& \multirow{2}{*}{Open-ended}& Librispeech~\citep{7178964} & 1000& \multirow{2}{*}{Acc}\\
 & && TIMIT~\citep{lamel1989speech} & 1000& \\
  \cmidrule{2-6}
 & \multirow{2}{*}{Dialogue generation}& \multirow{2}{*}{Open-ended}& DailyTalk~\citep{lee2023dailytalk} & 1000& \multirow{2}{*}{Acc}\\
 & && MELD~\citep{poria-etal-2019-meld}& 304 & \\
  \cmidrule{2-6}
 & \multirow{2}{*}{Hotword detection}& \multirow{2}{*}{Closed-form} & Librispeech~\citep{7178964} & 500 / 500 & \multirow{2}{*}{Acc}\\
 & && TIMIT~\citep{lamel1989speech} & 500 / 500 & \\
  \cmidrule{2-6}
 & \multirow{2}{*}{Speech identification}& \multirow{2}{*}{Closed-form} & Librispeech~\citep{7178964} & 500 / 500 & \multirow{2}{*}{F1/Acc} \\
 & && WSJ~\citep{paul-baker-1992-design} & 500 / 500 & \\
  \cmidrule{2-6}
 & \multirow{2}{*}{Keyword comparison} & \multirow{2}{*}{Closed-form} & AudioKeyword~\citep{mazumder2021multilingual}& 513 / 513 & \multirow{2}{*}{F1/Acc} \\
 & && SpeechCommands~\citep{warden2018speech} & 500 / 500 & \\
  \cmidrule{2-6}
 & Intent identification & Closed-form& FLUENT~\citep{lugosch19_interspeech}& 504 / 507 & F1/Acc\\
 \midrule
\multirow{9}{*}{Sound} & \multirow{2}{*}{Story generation} & \multirow{2}{*}{Open-ended}& ESC-50~\citep{piczak2015dataset}& 1000& \multirow{2}{*}{Acc}\\
 & && UrbanSound~\citep{Salamon:UrbanSound:ACMMM:14}& 1000& \\
 \cmidrule{2-6}
 & \multirow{2}{*}{Sound comparison} & \multirow{2}{*}{Closed-form} & ESC-50~\citep{piczak2015dataset}& 500 / 500 & \multirow{2}{*}{F1/Acc} \\
 & && UrbanSound~\citep{Salamon:UrbanSound:ACMMM:14}& 500 / 500 & \\
  \cmidrule{2-6}
 & \multirow{2}{*}{Caption retrieval}& \multirow{2}{*}{Closed-form} & Clotho~\citep{drossos2020clotho}& 506 / 494 & \multirow{2}{*}{Acc}\\
 & && AudioCaps~\citep{kim-etal-2019-audiocaps} & 528 / 472 & \\
  \cmidrule{2-6}
 & \multirow{2}{*}{Event retrieval}& \multirow{2}{*}{Closed-form} & ESC-50~\citep{piczak2015dataset}& 503 / 497 & \multirow{2}{*}{Acc}\\
 & && UrbanSound~\citep{Salamon:UrbanSound:ACMMM:14}& 503 / 497 & \\
  \cmidrule{2-6}
 & Event detection & Closed-form& AudioSet~\citep{7952261}& 500 / 500 & Acc \\
 \bottomrule
\end{tabular}
}
\caption{Data statistics of MAE benchmark. For closed-form tasks, class-wise number of samples are given.}
\label{tab:data-stats}
\end{table*}

\subsection{Speech Tasks}
For speech, we design two open-ended generation tasks from sentence and dialogue levels and four closed-form tasks from word and sentence levels.

\textbf{Speech comparison:}
This sentence-level, open-ended task requires ALLMs to identify content differences between pairs of speeches. 
Speech pairs are sourced from ASR datasets with timestamp-level transcription. 
We segment the speech into spoken words with timestamps and instruct an LLM to reconstruct a subset of these words into a new speech without semantic errors, such as omitting the adjective "expensive" in Fig.~\ref{fig:eval-overview}.
Then, we combine the original and reconstructed speech or two reconstructed speeches as an evaluation pair.
Note that ground truth transcriptions of two speeches can be obtained from the original labels, allowing us to utilize the LLM as an evaluator to label the ALLMs' responses based on the derived transcriptions.

\textbf{Dialogue response generation:} 
This dialogue-level, open-ended task involves generating a subsequent utterance from the examined ALLMs in a dialogue based on the first two utterances. 
Similarly, the LLM evaluator assesses the overall quality of the generated dialogue response with respect to the transcriptions of the preceding utterances.

\textbf{Hotword detection:} 
Hotword detection is defined as a word-level, close-ended task derived from existing ASR datasets. 
Given a pair of randomly sampled speeches, the ALLMs are asked to detect which speech contains a specific noun that appears only in one of them. 
This task evaluates the models' precision in recognizing and differentiating specific lexical items within speech contexts.

\textbf{Intent identification:} 
This task is defined as a sentence-level, close-ended task. We construct context speech pairs by randomly sampling speeches that either share the same intent or represent different intents from existing intent classification datasets. The ALLMs are tasked with determining whether the paired speeches belong to the same intent category, evaluating their ability to discern and categorize the underlying communicative purposes in speech data.

\textbf{Keyword comparison:} 
This task is a word-level, close-ended challenge derived from existing keyword-spotting datasets. We sample pairs of speeches that either share the same keywords or contain different ones. The ALLMs are asked to determine whether the keywords presented in the two speeches are identical.

\textbf{Speech identification:} 
Speech identification is a sentence-level, close-formed task involving context-positive speech pairs and negative pairs with distinct transcriptions.
The ALLMs are required to discern whether speech pairs have the same content.
Positive pairs are formed by adding different background noises (e.g., coughing and sneezing in Fig.~\ref{fig:eval-overview}) to the same speech, yielding identical content but different backgrounds. Negative pairs consist of randomly selected speeches.

\subsection{Sound Tasks}
For sound, we design one open-ended and four closed-form tasks to challenge the ALLM comprehension of global and local level audio intricacies.

\begin{figure*}[ht]
\centering
  \includegraphics[width=0.98\linewidth]{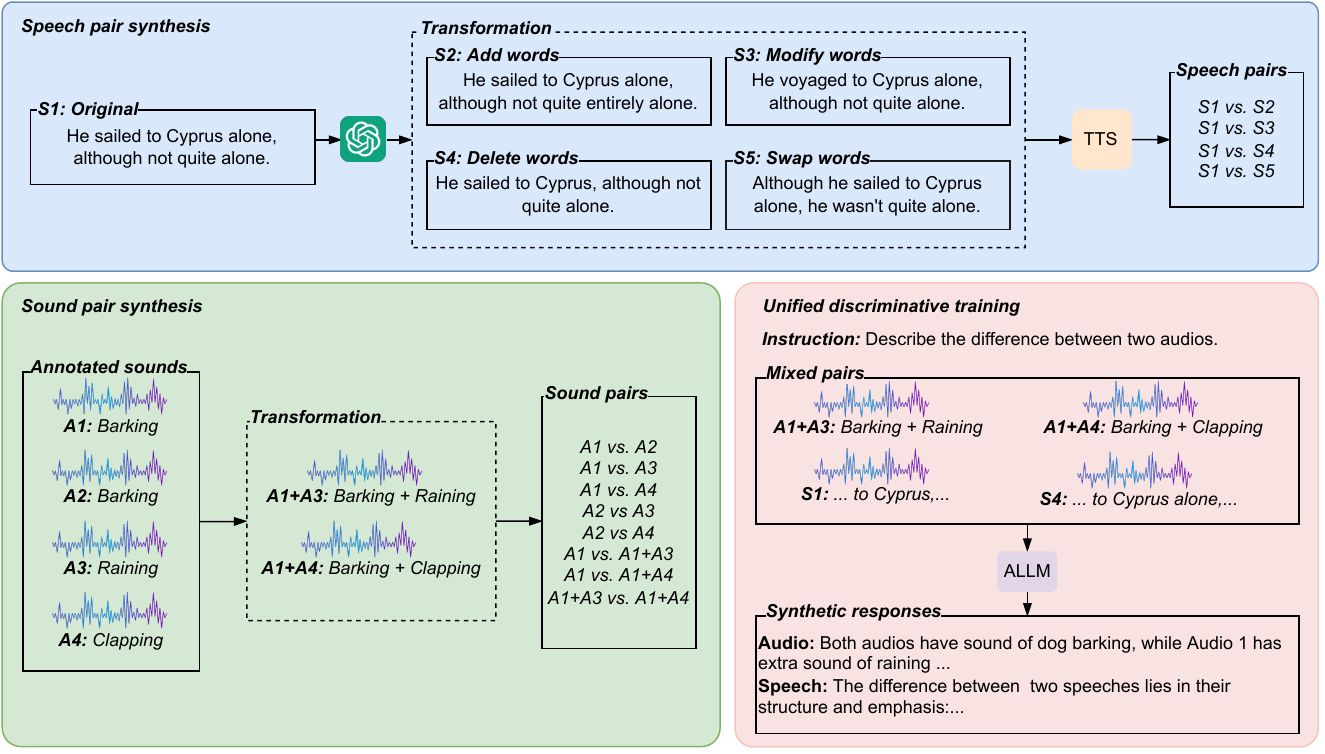} 
  \caption {MALLM training framework: speech/sound pair synthesis, and unified synthetic discriminative training.}
  \label{fig:train-framework}
\end{figure*}

\textbf{Story generation:}
This open-ended task tests the ALLMs' global understanding of audio by challenging them to integrate diverse sound events from multiple audios into a coherent narrative.
We randomly select two audio samples, each with a distinct event, and task the ALLMs to create a brief narrative incorporating both sounds.
The correctness of the story is contingent upon the inclusion of both specified sounds; any omission marks it as incorrect.
For instance, as shown in Fig.~\ref{fig:eval-overview}, a story is correct if it includes both the dog barking and bird chirping sounds.
We utilize the LLM evaluator to assess the completeness of the sound integration.

\textbf{Event detection:} 
Using multi-label sound event classification datasets, we create a task where ALLMs are asked to detect a specified sound in given audios containing multiple sounds, such as identifying if telephone ringing is present in Audio1 or Audio2 (Fig.~\ref{fig:eval-overview}).
This task demands precise understanding and interpretation of local audio details to identify specific sound events accurately.

\textbf{Sound comparison:} 
We randomly select two audio samples, which may either share the same sound event labels or differ.
ALLMs are then instructed to compare two audios to determine if these two audios belong to the same sound event category.
This requires the models to not only recognize the sound event in a single audio but also compare them for similarities and differences.

\textbf{Caption retrieval:} 
In this task, the ALLMs are presented with two audio samples and a corresponding audio caption. 
The models are required to determine which of the two audios aligns more closely with the given caption. 
This task assesses the ability of the ALLMs to understand audio content globally and analyze their similarities with descriptive text.

\textbf{Event retrieval:}
Similar to caption retrieval but simpler, ALLMs are asked to retrieve the audio sample based on one single specific sound event label each time, testing their ability to match labels with corresponding audio samples.

\section{MALLM}
In this section, we introduce the specifics of our proposed MALLM. Fig.~\ref{fig:train-framework} depicts our discriminative ALLM fine-tuning strategy. Specifically, MALLM is trained to describe the subtle distinctions between two similar audios, enhancing its capability to handle multi-audio scenarios.
To effectively tackle the challenging discriminative problem, the MALLM is expected to possess a deep comprehension of all input audios and infer both intra-audio and inter-audio relationships.
In this way, MALLM can be effectively augmented with reasoning abilities across multiple audios.
Furthermore, to improve the data efficiency and scalability, we propose the automatic construction of a synthetic training dataset without additional human intervention.
This dataset includes pairs of both synthetic speech and sound pairs with subtle differences, which are used to fine-tune MALLM, expanding its multi-audio processing capabilities.

\textbf{Speech pair synthesis:} 
To automate the construction of speech pairs, we start by randomly selecting a sentence from a text corpus. 
An advanced text-based LLM is then utilized to generate four variations of the sentence, each containing slight modifications. 
This is achieved by instructing the LLM with different prompts that direct it to add, delete, or modify a limited number of words in the original sentence and to alter its structure. 
Detailed prompts for this process are provided in APPX.~\ref{appx:data-synthesis-prompt}.
Following these text modifications, a robust pre-trained text-to-speech (TTS) model converts both the original and modified sentences into speech. This controlled synthesis process ensures that accurate ground-truth transcriptions are available for all generated speech samples. 
Ultimately, each original sentence is paired with its four modified counterparts, yielding four synthetic speech pairs with subtle differences per input sentence.

\begin{table*}[ht]
\centering
\resizebox{0.95\textwidth}{!}{ 
\begin{tabular}{lcccccc}
\toprule
\textbf{Model} & \multicolumn{2}{c}{\textbf{\begin{tabular}[c]{@{}c@{}}Speech comparison\\(Acc)\end{tabular}}} & \multicolumn{2}{c}{\textbf{\begin{tabular}[c]{@{}c@{}}Dialogue response generation\\(Acc)\end{tabular}}} & \multicolumn{2}{c}{\textbf{\begin{tabular}[c]{@{}c@{}}Speech identification\\(F1/Acc)\end{tabular}}}\\
\cmidrule(lr){2-3} \cmidrule(lr){4-5} \cmidrule(lr){6-7}
\textbf{}& \textbf{Librispeech}& \textbf{TIMIT}& \textbf{DailyTalk}& \textbf{MELD}& \textbf{Librispeech} & \textbf{WSJ} \\
\midrule
LTU-AS & 0.6 & 0.9 & 14.7& 11.5 & 66.7 / 50.0& 66.7 / 50.0\\
SALMONN-7B & 17.7& 17.5& 1.6 & 3.6& 66.7 / 50.0& 66.7 / 50.0\\
SALMONN-13B& 22.9& 27.9& 2.2 & 3.6& 68.5 / 54.1& 67.3 / 51.5\\
SpeechGPT & 1.2& 0.6& 1.9& 3.3& 2.0 / 50.5 & 1.6 / 50.2 \\
AnyGPT & 3.7& 2.9& 2.1& 3.0& 27.6 / 44.5& 32.5 / 48.8\\
Qwen-Audio & 11.6& 13.1& 20.4& 15.5 & 1.58 / 50.2& 0.00 / 50.0\\
\midrule
Gemini-Flash & 66.4 & 55.8 & 89.7 & 56.6 & 99.4 / 99.4& 99.6 / 99.6\\
Gemini-Pro & 60.5& 52.5& 79.0& 44.4 & 99.3 / 99.3& 99.2 / 99.2\\
\midrule
MALLM& \textbf{70.3} & \textbf{59.6} & \textbf{50.7} & \textbf{40.5}& \textbf{96.5 / 96.6} & \textbf{87.6 / 89.0} \\
\bottomrule
\toprule
 & \multicolumn{2}{c}{\textbf{\begin{tabular}[c]{@{}c@{}}Hotword detection\\(Acc)\end{tabular}}} & \multicolumn{2}{c}{\textbf{\begin{tabular}[c]{@{}c@{}}Keyword comparison\\(F1/Acc)\end{tabular}}}& \textbf{\begin{tabular}[c]{@{}c@{}}Intent identification\\(F1/Acc)\end{tabular}} & \textbf{\begin{tabular}[c]{@{}c@{}}Avg.\\(Acc)\end{tabular}} \\
 \cmidrule(lr){2-3} \cmidrule(lr){4-5} \cmidrule(lr){6-6}
 & \textbf{Librispeech}& \textbf{TIMIT}& \textbf{SpeechCommand}& \textbf{AudioKeyword}& \textbf{FLUENT}& \multicolumn{1}{l}{\textbf{}}\\
\midrule
LTU-AS & 49.8& 49.9& 52.6 / 49.2 & 64.0 / 48.3& 66.6 / 50.3& 34.1 \\
SALMONN-7B & 61.4& 55.7& 64.5 / 50.3 & 65.2 / 48.8& 66.8 / 50.2& 37 \\
SALMONN-13B& 51.9& 51.2& 0.4 / 50.0& 6.0 / 50.9 & 1.2 / 50.2 & 37.9 \\
SpeechGPT & 50.0 & 50.1 & 10.1 / 48.0& 11.2 / 48.7& 20.8 / 50.3& 32.3 \\
AnyGPT & 49.0 & 49.8 & 2.7 / 49.1 & 3.7 / 49.4 & 14.8 / 49.9& 32.0 \\
Qwen-Audio & 60.9& 63.9& 0.0 / 50.0& 0.0 / 50.0 & 66.8 / 50.2& 39.6 \\
\midrule
Gemini-Flash & 99.9 & 100.0& 91.7 / 92.3& 79.4 / 82.8& 80.2 / 75.5& 83.5 \\
Gemini-Pro & 99.7& 99.8& 94.9 / 95.0 & 89.36 / 90.16& 83.0 / 79.6& 81.7 \\
\midrule
MALLM& \textbf{80.0} & \textbf{86.8} & \textbf{88.2 / 88.7}& \textbf{78.03 / 80.90} & \textbf{75.7 / 68.6} & \textbf{73.8} \\
\bottomrule
\end{tabular}
}
\caption{Comparison of various ALLMs on MAE-Speech. The first block shows the open-source ALLMs, while the second block shows the proprietary ALLMs. The best performance among open-source ALLMs are bold.}
\label{tab:main-speech}
\end{table*}

\textbf{Audio pair synthesis:} 
To generate audio pairs with synthetic data, we start from individual audio recordings labeled with distinct sound events. 
For two individual audios containing different sound events, we directly mix the audios with balanced signal-to-noise ratios (SNR), resulting in a new audio file containing both sound events.
For instance, mixing an audio of a dog barking with an audio of rain produces a new synthetic audio featuring both barking and raining.
This mixing process ensures the ground truth sound event labels for the synthetic data are accurately maintained. Ultimately, various combinations of these mixed audios are generated to form audio pairs, as illustrated in Fig.~\ref{fig:train-framework}.

\textbf{Discriminative learning:} 
Following the acquisition of synthetic data pairs, we fine-tune the MALLM through inter-audio discriminative training tasks encompassing a mixture of speech and audio pairs with the instruction tuning loss.
As previously stated, we have access to ground truth labels for all synthetic data, facilitating the straightforward transformation of these labels into natural responses for training purposes.
For speech pairs, we utilize an advanced text-only LLM to generate differences between two speech transcriptions as the ground-truth responses.
Regarding audio pairs, we apply predefined rules to map annotated sound event labels to their corresponding responses (see APPX.~\ref{appx:train-data}).
To stabilize the training and prevent catastrophic forgetting, we also sample single audios from various tasks for joint training. 
Details regarding the MALLM training dataset are provided in APPX.~\ref{appx:train-data}.

\section{Experiment}
\subsection{Experiment Setup}
Throughout the experiments, we use GPT-4\footnote{gpt-4-turbo-2024-04-09} for data synthesis and downstream performance evaluation.
For the TTS module in speech pair synthesis, we utilize MMS-TTS~\citep{JMLR:v25:23-1318}.
We implement the MALLM based on Qwen-Audio~\citep{chu2023qwen}, incorporating an audio encoder, Whisper-large-v2~\citep{radford2023robust}, and a large language model, Qwen-7B~\citep{bai2023qwen}.
We use LoRA~\citep{hu2022lora} to fine-tune all modules of Qwen-Audio~\citep{chu2023qwen} except the audio encoder with a learning rate 5e-5 and batch size 16 for 5 epochs.
The training takes 48 GPU hours on Nvidia A100.

\begin{table*}[ht]
\centering
\resizebox{\textwidth}{!}{ 
\begin{tabular}{lcccccccccc}
\toprule
\textbf{Model} & \multicolumn{2}{c}{\textbf{\begin{tabular}[c]{@{}c@{}}Sound comparison\\(F1/Acc)\end{tabular}}} & \multicolumn{2}{c}{\textbf{\begin{tabular}[c]{@{}c@{}}Caption retrieval\\(Acc)\end{tabular}}} & \multicolumn{2}{c}{\textbf{\begin{tabular}[c]{@{}c@{}}Event retrieval\\(Acc)\end{tabular}}} & \textbf{\begin{tabular}[c]{@{}c@{}}Sound detection\\(Acc)\end{tabular}} & \multicolumn{2}{c}{\textbf{\begin{tabular}[c]{@{}c@{}}Story generation\\(Acc)\end{tabular}}} & \textbf{\begin{tabular}[c]{@{}c@{}}Avg.\\(Acc)\end{tabular}} \\
\cmidrule(lr){2-3} \cmidrule(lr){4-5} \cmidrule(lr){6-7} \cmidrule(lr){8-8} \cmidrule(lr){9-10}
\textbf{}& \textbf{ESC-50} & \textbf{UrbanSound} & \textbf{Clotho} & \textbf{AudioCaps}& \textbf{ESC-50} & \textbf{UrbanSound} & \textbf{AudioSet} & \textbf{ESC-50}& \textbf{UrbanSound} & \\
\midrule
LTU& 19.2 / 51.3 & 42.0 / 50.0 & 51.6& 51.4& 49.7& 50.3& 50.0& 25.7 & 27.5& 45.3\\
LTU-AS & 43.3 / 50.5 & 6.4 / 50.5& 50.5& 52.8& 49.4& 50.0& 50.0& 23.3 & 22.6& 44.4\\
Pengi & 3.1 / 50.0 & 2.0 / 49.8 & 50.3 & 52.7 & 50.3 & 50.3 & 50.0 & 0.9 & 0.3 & 39.4 \\
SALMONN-7B & 66.7 / 50.0& 66.7 / 50.0 & 53.3& 55.6& 50.8& 50.8& 50.7& 6.2& 13.4& 42.3\\
SALMONN-13B& 3.9 / 50.5 & 10.3 / 51.0 & 51.3& 50.4& 54.8& 53.1& 53.2& 25.3 & 47.5& 48.6\\
NextGPT-7B & 4.2 / 50.3 & 2.4 / 50.1& 50.4& 52.2& 50.4& 50.3& 50.0& 1.0& 0.8 & 39.5\\
PandaGPT-7B& 66.7 / 50.0 & 66.7 / 50.0 & 51.1& 54.9& 51.3& 52.0& 52.1& 4.3& 3.5 & 41.0\\
PandaGPT-13B & 45.5 / 48.7 & 44.6 / 52.7 & 50.8& 49.1& 51.1& 51.0& 52.8& 2.6& 1.7 & 40.1\\
XInstructBLIP-7B & 47.3 / 51.0 & 44.6 / 51.7 & 50.8& 52.8& 49.7& 50.3& 50.0& 3.4& 3.9 & 40.4\\
XInstructBLIP-13B & 2.7 / 50.1 & 9.7 / 49.5 & 50.8 & 53.2 & 50.0 & 49.5 & 49.7 & 7.0 & 4.5 & 40.5 \\
Qwen-Audio & 0.4 / 50.1& 17.7 / 50.0 & 58.8& 55.2& 83.1& 74.3& 68.5& 30.6 & 42.0& 57.0\\
\midrule
MALLM& \textbf{80.7 / 76.6}& \textbf{69.7 / 65.7}& \textbf{81.3} & \textbf{65.0} & \textbf{95.5} & \textbf{80.4} & \textbf{72.0} & \textbf{59.2}& \textbf{70.8} & \textbf{74.1} \\
\bottomrule
\end{tabular}
}
\caption{The performance of various ALLMs on MAE-Sound. The best performance among ALLMs are bold.}
\label{tab:main-audio}
\end{table*}

\subsection{Examined Models}
Our evaluation encompasses a diverse range of ALLMs on the MAE benchmark. 
For speech-related tasks, we assess LTU-AS~\citep{gong2023joint}, SALMONN~\citep{tang2024salmonn}, Qwen-Audio~\citep{chu2023qwen}, SpeechGPT~\citep{zhang-etal-2023-speechgpt} and AnyGPT~\citep{zhan2024anygpt}. 
In the domain of sound processing, we evaluate LTU~\citep{gong2024listen}, LTU-AS~\citep{gong2023joint}, SALMONN~\citep{tang2024salmonn}, Qwen-Audio~\citep{chu2023qwen}, NextGPT~\citep{wu2023next}, PandaGPT~\citep{su-etal-2023-pandagpt}, X-InstructBLIP~\citep{panagopoulou2023x} and Pengi~\citep{deshmukh2023pengi}. 
In addition, we include advanced proprietary models, Gemini-Flash\footnote{gemini-1.5-flash-preview-0514} and Gemini-Pro\footnote{gemini-1.5-pro-preview-0514}, as the performance upper bound of speech tasks, since they cannot handle the sound tasks. 
Note that the examined ALLMs sometimes fail to follow the instructions, leading to the absence of predicted answers for closed-form tasks. In that case, we default the predicted answer to "No" or "Audio1", depending on the task context. Since MAE is class-balanced, this strategy effectively equates to a random guessing strategy.

\subsection{MAE Results}
\textbf{MAE-Speech:} 
The evaluation results for various ALLMs on the MAE-Speech benchmark (Tab.~\ref{tab:main-speech}) show that all open-source ALLMs struggle with multi-audio scenarios across all tasks.
The highest-performing model, Qwen-Audio, merely achieves an average accuracy of 39.6\%. 
In closed-form tasks, these models tend to collapse to consistently give the same answer for all queries, leading to approximately 50\% accuracy.
For example, Qwen-Audio always answers "Yes" for intent identification.
Furthermore, models like LTU-AS frequently fail to adhere to the provided instructions, resulting in extremely poor performance, particularly in open-ended tasks. 
This issue is hypothesized to stem from insufficient supervised fine-tuning data.

In contrast, the proprietary Gemini series models demonstrate exceptional performance across the MAE benchmark, reaching nearly 100\% accuracy in simple tasks like hotword detection.
This underscores the significant performance disparity between current open-source ALLMs and their proprietary counterparts.
Notably, the Gemini technical reports do suggest the multi-audio processing abilities of Gemini models and examine a multi-audio in-context learning scenario~\citep{reid2024gemini}. 
Therefore, it’s likely that Gemini models are specially optimized with multi-audio data.

Our newly developed MALLM significantly surpasses all open-source models, posting an average accuracy of 73.8\%. 
It is important to mention that the MALLM's tuning data does not encompass dialogue or intent-related data. 
Nevertheless, we observe substantial improvements in these areas, suggesting that our discriminative training strategy effectively enhances the general multi-audio processing capabilities of ALLMs.
Overall, our MALLM substantially narrows the gap between open-source and proprietary models, even surpassing proprietary models in tasks like speech comparison.

\textbf{MAE-Sound:} 
The evaluation results for various ALLMs on the MAE-Sound benchmark, shown in Tab.~\ref{tab:main-audio}, reveal that most ALLMs struggle with multi-audio contexts similar to the findings from MAE-Speech.
Among all ALLMs, Qwen-Audio achieves the best performance, particularly in event retrieval and sound detection tasks.
Despite this, its overall performance is still deemed unsatisfactory.
Our proposed MALLM significantly outperforms Qwen-Audio and all other models, achieving an average accuracy of 74.3\% across all tasks.
This superior performance again underscores the efficacy of our discriminative learning approach tailored for multi-audio contexts in large audio language modeling.
Overall, MALLM demonstrates a marked improvement over existing models in multi-audio processing across both speech and sound domains, which suggests promising directions for further research and development.

\subsection{Discussion}
\textbf{Single-audio performance:} 
Alongside multi-audio processing, we also evaluate the single-audio processing ability of the MALLM compared to the backbone model, Qwen-Audio, with results for ASR and event classification in Tab.~\ref{tab:single-results}.
Our MALLM not only demonstrates superior multi-audio understanding but also maintains competitive, sometimes superior, single-audio performance.
These findings underscore the MALLM’s broad applicability and validate the effectiveness of our synthetic training schema, which equips the model to handle a wide range of scenarios.

\begin{table}[ht]
\centering
\resizebox{\columnwidth}{!}{ 
\begin{tabular}{lcccc}
\toprule
\textbf{Model}& \multicolumn{2}{c}{\begin{tabular}[c]{@{}c@{}}\textbf{Librispeech ASR}\\\textbf{(WER)}\end{tabular}} & \multicolumn{2}{c}{\begin{tabular}[c]{@{}c@{}}\textbf{Event Classification}\\\textbf{(Acc)}\end{tabular}} \\
\cmidrule(lr){2-3} \cmidrule(lr){4-5}
 & \textbf{dev-clean}& \textbf{test-clean}& \textbf{ESC-50} & \textbf{UrbanSound} \\
\midrule
Qwen-Audio & \textbf{2.0}& 2.3 & 70.0 & \textbf{74.9} \\
MALLM& \textbf{2.0}& \textbf{2.2} & \textbf{77.2} & 73.5 \\
\bottomrule
\end{tabular}
}
\caption{Comparison of single-audio performance between Qwen-Audio and our MALLM.}
\label{tab:single-results}
\end{table}

\textbf{Case study:}
In addition to quantitative results, we show a case study in Tab.~\ref{tab:case_study} to illustrate the quality of generated responses from different ALLMs. 
We leave more case studies in APPX.~\ref{appx:case-study}.
In the example, the models need to generate a story with sounds of both keyboard typing and water dropping.
While both generate coherent narratives, Qwen-Audio focuses only on keyboard typing, neglecting the water sound.
On the contrary, MALLM incorporates both sounds and demonstrates a more comprehensive understanding of the task requirements, demonstrating its superior capabilities in handling complex multi-audio inputs compared to other models.

\begin{table}[ht]
    \centering
    \footnotesize
    \begin{tabularx}{\columnwidth}{lX}
    \toprule
    \textbf{Model} & \textbf{Response} \\
    \midrule
    Qwen-Audio & In a dimly lit room, a person sat at a desk typing furiously on a \textbf{keyboard}. The sound of the keys clacking against the keys filled the room. Suddenly, the person stopped typing and listened to the sound of their own typing. It was a soothing, calming sound that brought them a sense of peace. They closed their eyes and took a deep breath, letting the sound wash over them. They felt a sense of accomplishment and satisfaction, knowing that they had just typed out a masterpiece. \\
    \midrule
    MALLM & In a quiet, dimly lit room, there was a constant sound of typing on a \textbf{keyboard}. As the sound filled the air, drops of \textbf{water} began to fall from a faucet. The rhythmic clicking of the \textbf{keyboard} mixed with the rhythmic splashing of the \textbf{water}, creating a unique and soothing atmosphere. The combination of sounds brought a sense of calm and focus to the environment. The \textbf{keyboard} clicking was steady and consistent, much like the dripping \textbf{water}, and both sounds seemed to blend together, creating a harmonious and tranquil scene. \\
    \bottomrule
    \end{tabularx}
    \caption{Case study: ALLMs are instructed to generate a story containing the sound of water and keyboard. The occurrence of 'keyboard' and 'water' in the model response is in bold.}
    \label{tab:case_study}
\end{table}

\textbf{Human validation of automatic metrics:}
For open-ended tasks in MAE, we employ an LLM evaluator to automatically assess the correctness of responses from ALLMs. 
To validate the reliability of this evaluator, we manually annotate responses from Qwen-Audio and MALLM, which exhibit the best performance among open-source models. 
We randomly selected 50 responses from each model across six datasets within three tasks: speech comparison, dialogue response generation, and story generation.
For each sample, we ask three annotators to annotate the binary label of the ALLM response. 

We observe a high inter-annotator agreement of 0.81 for the story generation task, and a moderate inter-annotator agreement of 0.39 and 0.38 for dialogue response generation and speech comparison respectively, which validates the reliability of our human study.
The accuracy computed from LLM annotated labels and human annotated ones are summarized in Tab.~\ref{tab:human-results}.
Across all tasks, particularly in story generation, the LLM evaluator shows a high correlation with human judgments, affirming its reliability as a tool in our MAE benchmark. 
Note that the evaluation of speech comparison and dialogue response are subjective and challenging.
Even different human annotators exhibit different perceptions, as highlighted in the inter-annotator agreement.
Therefore, despite the slightly lower accuracy, the LLM evaluator still serves as a robust tool for providing reliable assessments.

\begin{table}[ht]
\centering
\resizebox{\columnwidth}{!}{ 
\begin{tabular}{ccc}
\toprule
\textbf{Speech comparison} & \textbf{Dialogue generation} &\textbf{Story generation} \\
\midrule
71.5   & 78.5   & 88.5 \\
\bottomrule
\end{tabular}
}
\caption{The accuracy of LLM evaluator against human annotators.}
\label{tab:human-results}
\end{table}

\section{Conclusion}
In this paper, we introduce the first multi-audio evaluation benchmark, MAE, to examine the multi-audio processing ability of ALLMs.
Extensive evaluations across 15 ALLMs reveal generally unsatisfactory performance among current open-source models in handling multi-audio scenarios. 
To this end, we develop a simple and effective discriminative training framework that leverages synthetic data.
The resulting model, MALLM, significantly surpasses all existing open-source ALLMs without incurring additional costs for human annotation.
This work lays a robust foundation for future research aimed at enhancing ALLM capabilities in multi-audio processing. 
Future work includes introducing more complex multi-audio tasks that involve more than two audios and expanding the MALLM training dataset to cover more scenarios.

\section*{Limitations} In this paper, we propose the first benchmark to evaluate the multi-audio analysis ability of ALLMs. We also develop a novel MALLM, which outperforms existing ALLMs on multi-audio processing. Yet, there are several limitations to this work. 
Firstly, considering the large performance gap between open-source and proprietary models, we deliberately design the tasks in MAE with a simple nature to obtain meaningful results from open-source ALLMs. Therefore, while MAE poses large challenges to various open-source ALLMs, some tasks are too simple for proprietary ALLMs like Gemini. In future work, we plan to explore more complex tasks, such as speech recognition~\citep{gao2024transferable,yue2024adapting}, compositional reasoning~\citep{ghosh2024compa}, speech synthesis~\citep{gao2024ttslow} and lyrics transcription~\citep{gao2023self}.
Secondly, the proposed MAE benchmark currently covers English data. Extending the MAE to the multilingual scenario is also an important future direction to ensure that it's comprehensive and applicable across diverse linguistic contexts.
Thirdly, the proposed MALLM is currently trained on a relatively small scale of synthetic data due to computational resource constraints. We aim to further enhance the MALLM with large-scale, diverse training data, enabling its application to more challenging domains like singing \cite{gupta2019nus} and music \cite{gao2023polyscriber} in future work.

\section*{Acknowledgments}
This work is supported in part by Shenzhen Science and Technology Program (Grant No. ZDSYS20230626091302006) and supported in part by Shenzhen Science and Technology Research Fund (Fundamental Research Key Project Grant No. JCYJ20220818103001002). This work is also supported in part by the European Commission through Project ASTOUND (101071191 — HORIZON-EIC-2021- PATHFINDERCHALLENGES-01), and by project BEWORD (PID2021-126061OB-C43) funded by MCIN/AEI/10.13039/501100011033 and, as appropriate, by “ERDF A way of making Europe”, by the “European Union”.
This research is supported by National Research Foundation Singapore under its AI Singapore Programme (Award Grant No: AISG2-100E-2022-102).

\bibliography{bib/anthology_1,bib/anthology_2,bib/custom}

\appendix

\section{Evaluation Prompt}
\label{appx:eval-prompt}
We utilize GPT-4 to automatically give binary quality labels for open-ended generation tasks. To get more stable and consistent predictions from GPT-4, we produce 3 predictions from GPT-4 through repeatedly sampling and take the majority vote as the final prediction.
The GPT-4 automatic evaluation prompts for different tasks are listed:
\begin{itemize}
    \item \textbf{Speech comparison:}
    \begin{tcolorbox}
    \small
\texttt{Given the transcriptions of two speeches.}

\texttt{Speech 1: [[Speech 1]]}

\texttt{Speech 2: [[Speech 2]]}

\texttt{Determine whether the below response correctly captures the difference between speech 1 and speech 2.}

\texttt{[[Response]]}

\texttt{Your answer should be a single "Yes" or "No". Do not output anything else.}
    \end{tcolorbox}
    \item \textbf{Dialogue response generation:}
        \begin{tcolorbox}
    \small
\texttt{Given the first two utterances from a conversation between speaker 1 and speaker 2.}

\texttt{Speaker 1: [[Speech 1]]}

\texttt{Speaker 2: [[Speech 2]]}

\texttt{For the question:}

\texttt{What would be the possible utterance 3 from speaker 1 as a response to utterance 2 from speaker 2?}

\texttt{Determine whether the below response correctly answers the question.}

\texttt{[[Response]]}

\texttt{Your answer should be a single "Yes" or "No". Do not output anything else.}
    \end{tcolorbox}
    \item \textbf{Story generation:}
        \begin{tcolorbox}
    \small
\texttt{Given the story:}

\texttt{[[Response]]}

\texttt{Determine whether the below two sounds are presented in the story.}

\texttt{Sound1: [[Speech 1]]}

\texttt{Sound2: [[Speech 2]]}

\texttt{Answer "Yes" if both sounds are presented in the story. Otherwise, answer "No".}

\texttt{Your answer should be a single "Yes" or "No". Do not output anything else.}
    \end{tcolorbox}
    
\end{itemize}

For all tasks, "No" indicates that the ALLM's response is inappropriate, while "Yes" indicates that the response is suitable.

\section{MALLM Data}
\label{appx:train-data}
In Tab.~\ref{tab:audio-reponse-format}, we provide the example responses of possible input audio pair combinations.
In the first four cases, the ALLM is asked to describe the differences between two audios.
For the last cases, where there's a common sound event occur in both audios, we also ask the ALLM to answer the common sound event in two audios.

\begin{table}[ht]
    \centering
    \small
    \begin{tabularx}{\columnwidth}{lX}
    \toprule
    \textbf{Audio Pair} & \textbf{Example Response} \\
    \midrule
    A1 vs. A2 & Both audios have the same sound event of [barking]. \\
    A1 vs. A3 & Audio1 has sound of [barking], while Audio2 has sound of [raining]. \\
    A1 vs. A1+A3 & Both audios have sound of [barking], while Audio2 has additional sound of [raining]. \\
    A1+A3 vs. A1+A4 & Both audios have sound of [barking], while Audio1 has additional sound of [raining] and Audio2 has additional sound of [clapping]. \\
    A1+A3 vs. A1+A4 & The common sound event in two audios is [barking]. \\
    \bottomrule
    \end{tabularx}
    \caption{Example response output for audio discriminative training. The A1,A2,A3,A4 refer to audio clips in Fig.~\ref{fig:train-framework}. The [x] is the ground truth sound event label.}
    \label{tab:audio-reponse-format}
\end{table}

The data statistics of the final MALLM training data is listed in Tab.~\ref{tab:train-data-stats}.
We use the test split of data source for all tasks in MAE benchmark, and use the train split of data source to construct our training data. 
For human validation of the LLM evaluator, we provide the human annotators with the same instructions given to the LLM evaluator.
In addition, both raw audios and ground truth labels are provided to the human annotators.

\begin{table}[ht]
\centering
\resizebox{\columnwidth}{!}{ 
\begin{tabular}{lcc}
\toprule
\textbf{Data}& \textbf{Source} & \textbf{\# Samples} \\
\midrule
Speech pairs& BookCorpus~\citep{Zhu_2015_ICCV} & 24K \\
Sound pairs& VGGSound~\citep{9053174}& 26.9K \\
\midrule
Speech recognition & Librispeech~\citep{7178964} & 28.5K \\
Sound classification & VGGSound~\citep{9053174}& 20.7K \\
Audio caption& AudioCaps~\citep{kim-etal-2019-audiocaps} & 5K\\
\midrule
Total& & 105.1K \\
\bottomrule
\end{tabular}
}
\caption{MALLM training data statistics. The first block shows the synthetic multi-audio data, and the second block shows the single-audio data.}
\label{tab:train-data-stats}
\end{table}

\section{Data Synthesis Prompt}
\label{appx:data-synthesis-prompt}
We utilize LLM to obtain sentence pairs with the below four different prompts:
\begin{itemize}
\item \textbf{Add word:}
\begin{tcolorbox}
\small
\texttt{Given the below sentence:}

\texttt{[[Sentence]]}
\newline

\texttt{You need to modify this sentence to a new one by adding several words. The number of added words should be less than 3.}

\texttt{Note that adding is the only operation you can do.}

\texttt{The new sentence needs to be semantically correct without error. Try to be creative.}

\texttt{Your output should be the new sentence only without anything else.}

\texttt{If you cannot generate a new sentence meeting the above criteria (less than 3 added words and semantically correctness), output a single word "None".}
\end{tcolorbox}
\item \textbf{Delete word:}
\begin{tcolorbox}
\small
\texttt{Given the below sentence:}

\texttt{[[Sentence]]}
\newline

\texttt{You need to modify this sentence to a new one by deleting several words. The number of deleted words should be less than 3.}

\texttt{Note that deletion is the only operation you can do.}

\texttt{The new sentence needs to be semantically correct without error. Try to be creative.}

\texttt{Your output should be the new sentence only without anything else.}

\texttt{If you cannot generate a new sentence meeting the above criteria (less than 3 deleted words and semantically correctness), output a single word "None".}
\end{tcolorbox}
\item \textbf{Modify word:}
\begin{tcolorbox}
\small
\texttt{Given the below sentence:}

\texttt{[[Sentence]]}
\newline

\texttt{You need to modify this sentence to a new one. The edit distance between the generated sentences and the old one needs to be less than 5.}

\texttt{The new sentence needs to be semantically correct without error. Try to be creative.}

\texttt{Your output should be the new sentence only without anything else.}

\texttt{If it's impossible to have a new sentence meeting the above criteria (edit distance and semantically correctness), output "None".}
\end{tcolorbox}

\item \textbf{Change structure:}
\begin{tcolorbox}
\small
\texttt{Given the below sentence:}

\texttt{[[Sentence]]}
\newline

\texttt{You need to derive a new sentence by altering the structure of the given sentence.}

\texttt{Remember to keep the wording usage unchanged.}

\texttt{The difference between two sentences need to be less than 5 words.}

\texttt{The new sentence needs to be semantically correct without error. Try to be creative.}

\texttt{Your output should be the new sentence only without anything else.}

\texttt{If it's impossible to have a new sentence meeting the above criteria (less than 5 words difference and semantically correctness), output "None".}
\end{tcolorbox}

\end{itemize}

\section{Case Study}
\label{appx:case-study}
We compare the generated responses from Qwen-Audio and MALLM on dialogue response generation and speech comparison in Tab.~\ref{tab:case_study_extended}. 
Similar to our findings from Tab.~\ref{tab:case_study}, the Qwen-Audio fails to combine the context from two input speeches to infer the correct response. 
On the contrary, the proposed MALLM can handle the multi-audio scenarios.

\begin{table}[ht]
\centering
\footnotesize
\begin{tabularx}{\columnwidth}{lX}
\toprule
\multicolumn{2}{c}{\textit{DailyTalk Response Generation}} \\
\midrule
Speech1 & I want to buy some jewellery.   \\
Speech2 & What kind of jewellery do you like to have? \\
\midrule
Qwen-Audio& I want to buy some jewelry  \\
MALLM   & I like to have a diamond ring.\\
\bottomrule
\toprule
\multicolumn{2}{c}{\textit{LibriSpeech Speech Comparison}}\\
\midrule
Speech1 & creatures attacked each other\\
Speech2 & huge creatures attacked each other \\
\midrule
Qwen-Audio& The content difference between speech 1 and speech 2 is that in speech 1, "creature" is mentioned, in speech 2, "huge" is mentioned. \\
MALLM & The difference between Speech 1 and Speech 2 lies in the adjective used to describe the creatures. In Speech 1, the creatures are simply mentioned without any specific description, implying they could be of any size or type. In Speech 2, the adjective "huge" is used to specifically describe the creatures, suggesting that they are very large. This additional detail in Speech 2 emphasizes the size of the creatures, potentially altering the reader's perception of the scene's intensity or the creatures' capabilities. \\
\bottomrule
\end{tabularx}
\caption{Case study on dialogue response generation and speech comparison.}
\label{tab:case_study_extended}
\end{table}

\end{document}